\newcommand{\etal}{{\it et al.}}

\documentclass{ws-ijmpa}

\begin{document}

\markboth{R. Sia} {Evidence for $B_s^{(*)}$ Production at the
$\Upsilon$(5S)}

\catchline{}{}{}{}{}
%
%%%%%%%%%%%%%%%%%%%%%%%%%%%%%%%%%%%%%%%%%%%%%%%%%%%%%%%%%%%%%%%%%%%%

\title{\large Evidence for $B_s^{(*)}$
Production at the $\Upsilon$(5S)}

\author{\footnotesize R. Sia \\ Representing the CLEO Collaboration}

\address{Department of Physics, Syracuse University, Syracuse, New York 13244\\
E-mail: rsia@phy.syr.edu}

\maketitle

%\pub{Received (Day Month Year)}{Revised (Day Month Year)}

\begin{abstract}
  Based on data collected by the CLEO III detector at CESR, we started a
  series of investigations to open the mysteries of the $\Upsilon$(5S) resonance decay
  properties. $B_s$
  mesons are expected to decay predominantly into $D_s$ meson, while
  the lighter $B$ mesons decay into $D_s$ only about 10\% of the time.
  We exploit this difference to present the first evidence of a substantial
  production of $B_s$ mesons at the $\Upsilon$(5S) resonance. We make
  here a preliminary model dependent
  estimate of the ratio of $B_s^{(*)}\overline{B}_s^{(*)}$ to the
  total $b\overline{b}$ quark pair production at the $\Upsilon$(5S)
  energy to be $(21\pm 3 \pm 9)$\%.
  \end{abstract}
%  \pacs{13.20.He} \maketitle
%\keywords{Keyword1; keyword2; keyword3.}
\section{Introduction}
 An enhancement in the total $e^+e^-$
annihilation cross-section into hadrons was discovered at CESR
long ago\cite{CLEOIImeasurement} ,\cite{CUSBhyperfine}
,\cite{CUSBmasses} at 10.865$\pm$0.008 GeV. This effect was named
the $\Upsilon$(5S) resonance. Potential models\cite{paper} predict
the different relative decay rates of the $\Upsilon$(5S) into
combinations of $B^{(*)}\overline{B}^{(*)}$ and $B_s^{(*)}
\overline{B}_s^{(*)}$. In a simple spectator model the $B_s$
decays into the $D_s$ nearly all the time. Since the $B\to D_s X$
branching ratio has already been measured to be $(10.5\pm 2.6\pm
2.5)\%$,\cite{PDG} we expect a large difference between the $D_s$
yields at the $\Upsilon$(5S) and the $\Upsilon$(4S) that we
examine in this paper and that can lead to an estimate of the size
of the $B_s^{(*)}\overline{B}_s^{(*)}$ component at the
$\Upsilon$(5S). By the $\Upsilon$(5S) here, we mean the production
of any $B$ meson species including $B_u$, $B_d$ and $B_s$ that
occurs at an $e^+e^-$ center-of-mass energy of 10.87 GeV.

\section{Analysis Technique}

The data sample we used for this analysis consists of 0.42
  fb$^{-1}$ of data taken on the $\Upsilon$(5S) resonance, 6.34
  fb$^{-1}$ of data collected on the $\Upsilon(4S$) and 2.32 fb$^{-1}$
  of data taken in the continuum below the $\Upsilon$(4S).
  We reconstruct $D_s$ mesons via the
$D_s^+\rightarrow\phi\pi^+$ decay mode using the $\phi\to K^+K^-$
channel (Charge conjugate decays are implied throughout this
paper). Details about track and event selection criteria are
available elsewhere.\cite{paper} We consider $D_s$ candidates
having a momentum less than or equal to half of the beam energy,
the limit from a $B$ decay. To remove at first order differences
between continuum data taken just below the $\Upsilon$(4S), at the
$\Upsilon$(4S) and at the $\Upsilon$(5S) we choose to work with
the variable $x$ which is the $D_s$ momentum divided by the beam
energy. The number of signal events is determined by subtracting
the scaled continuum data below the $\Upsilon$(4S) from the
$\Upsilon$(4S) and from the $\Upsilon$(5S) data.\cite{paper} The
$x$ dependent $D_s$ detection efficiencies of the two datasets are
consistent with each other and have values around 30\%.

\section{$D_s$ production rates at the $\Upsilon$(4S) and the
$\Upsilon$(5S)}

\begin{figure}[htbp]
 \centerline{\epsfig{figure=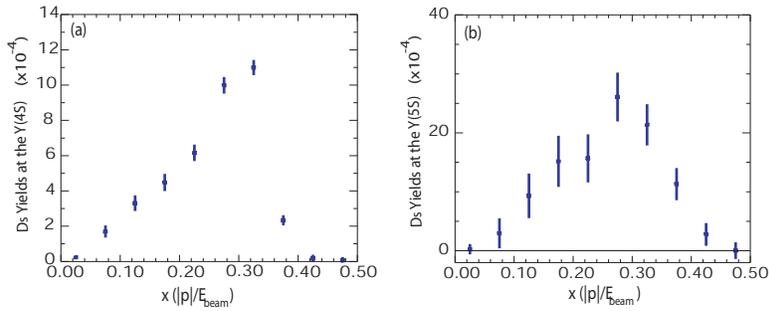,height=1.6in, width=4in}}
  \caption{$D_s$ yields vs x from: (a) the $\Upsilon$(4S) decays, (b) the $\Upsilon$(5S)
  decays. Both plots are continuum subtracted, efficiency corrected, and normalized to the number of
  resonance events (Preliminary). The partial production rates at the different energies
vs. $x$ are available.$^4$}\label{Dsyield}
\end{figure}

In Fig.~\ref{Dsyield}(a) and \ref{Dsyield}(b), we show the $x$
distribution of the inclusive $D_s$ yields from the $\Upsilon$(4S)
and the $\Upsilon$(5S) decays respectively.
%, continuum subtracted, efficiency
%corrected, and normalized to the number of resonance events.
Using $6,420,910\pm 5,738\pm 240,542$ of $\Upsilon$(4S) resonance
events and $131,396\pm 810 \pm 26,546$ of $\Upsilon$(5S) resonance
events above the four-flavor continuum, we find
${\cal{B}}(\Upsilon(4S)\to D_sX)\cdot{\cal{B}}(D_s\to\phi\pi)
=(8.0\pm 0.3 \pm 0.4)\cdot{10^{-3}}$ and
${\cal{B}}(\Upsilon(5S)\to
D_sX)\cdot{\cal{B}}(D_s\to\phi\pi)=(20\pm 2 \pm 4)\cdot{10^{-3}}$,
thus demonstrating a much larger production of $D_s$ at the
$\Upsilon$(5S) energy than at the $\Upsilon$(4S),
i.~e.~${\cal{B}}(\Upsilon(5S)\to D_s X)/{{\cal{B}}(\Upsilon(4S)\to
D_s X)}=2.5\pm 0.3\pm 0.6$. Using ${\cal{B}}(D_s\to\phi\pi^+)= (
3.6 \pm 0.9 )\%$,\cite{PDG} we measure ${\cal{B}}(\Upsilon(4S)\to
D_sX)$ to be $(22.3\pm 0.7\pm 5.7)\%$ hence ${\cal{B}}(B\to
D_sX)=(11.1\pm0.4\pm2.9)\%$ which is in a good agreement with the
PDG\cite{PDG} value of $(10.5\pm 2.6 \pm 2.5)\%$. In addition, we
find ${\cal{B}}(\Upsilon(5S)\to D_sX)=(55.0\pm 5.2 \pm 17.8)\%$.

\section{$B_s$ production at the $\Upsilon$(5S)}
\begin{figure}[htbp]
 \centerline{\epsfig{figure=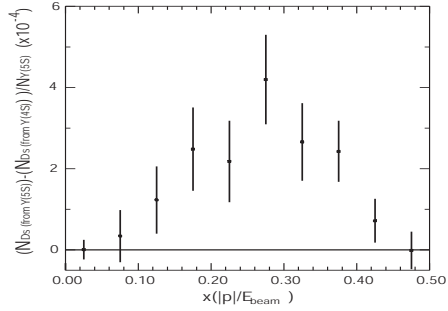,height=1.6in, width=2.3in}}
  \caption{The enhancement of $D_s$ yields at the
$\Upsilon$(5S) vs $x$ (no efficiency correction,
Preliminary).}\label{enhancement}
\end{figure}
\begin{figure}[htbp]
 \centerline{\epsfig{figure=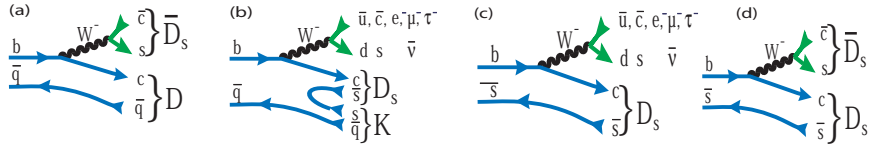,height=0.8in, width=4.5in}}
  \caption{\label{BtoDs} Dominant decay diagrams for a $B$ and $B_s$ meson into $D_s$ mesons
  ($q$ can be here either a $u$ or a $d$ quark).}
\end{figure}

In Fig.~\ref{enhancement}, we show the $D_s$ meson yields vs. $x$
from the $\Upsilon$(5S) with the $\Upsilon$(4S) spectrum
subtracted. The spectrum shows a significant excess of $D_s$ at
the $\Upsilon$(5S), which is a significant evidence for $B_s$
production at the $\Upsilon$(5S). The branching ratio we measure
${\cal{B}}(B\to D_sX)$ comes in fact either from the the $W^-\to
\overline {c}s$ or from the $b\to c$ piece if it manages to create
an $s\overline{s}$ pair through fragmentation as shown in
Fig.~\ref{BtoDs}(a) and \ref{BtoDs}(b) respectively.
Similarly, the production of $D_s$ mesons from $B_s$ decay arises
from two dominant processes as well. Fig.~\ref{BtoDs}(c) and
\ref{BtoDs}(d) show a large, possible greater than $100\%$ $D_s$
rate; here the primary $b\to c$ transition has the charm quark
pairing with the spectator anti-strange quark. $D_s$ can also be
produced from the upper vertex in Fig.~\ref{BtoDs}(c) when the
$W^-\to \overline {c}s$ and these two quarks form a color singlet
pair. The chances of this occurring should be similar to the
chance of getting an upper-vertex $D_s$ in $B$ decay
(Fig.~\ref{BtoDs}(a)). However, It is possible that some $D_s$ are
lost from these processes since any $c\overline{s}$ pairs in
Fig.~\ref{BtoDs}(c) and \ref{BtoDs}(d) could fragment into a kaon
with a $D$ particle instead of a $D_s$ by producing a
$u\overline{u}$ or $d\overline{d}$. These considerations lead to a
model dependent estimate of ${\cal{B}}(B_s\to D_sX)=(92\pm11)\%$.
Based on this estimate and using the measured product branching
fractions ${\cal{B}}(\Upsilon(5S)\to D_s
X)\cdot{\cal{B}}(D_s\to\phi\pi^+)$ and ${\cal{B}}(\Upsilon(4S)\to
D_s X)\cdot{\cal{B}}(D_s\to\phi\pi^+)$, we find the ratio of
$B_s^{(*)}\overline{B}_s^{(*)}$ to the total $b\overline{b}$ quark
pair production at the $\Upsilon$(5S) energy\\

%\begin{equation}
~~~~~~~~~~~~~~~~~~~~~~${\cal{B}}(\Upsilon(5S)\to
B_s^{(*)}\overline{B}_s^{(*)})=(21\pm 3 \pm 9 )\%~$
\\
%\end{equation}

\indent I gratefully acknowledge my advisor Professor S. Stone for
all his guidance and help.

\end{document}